\documentclass[
reprint,
superscriptaddress,
nofootinbib,
nobibnotes,
amsmath,amssymb,
aps,
prc,
]{revtex4-1}

\usepackage{graphicx}
\usepackage{dcolumn}
\usepackage{bm}
\usepackage{hyperref}
\usepackage{xcolor}
\usepackage{slashed}
\usepackage{ulem}
\usepackage{tabularx}
\usepackage{verbatim}

\usepackage{subcaption}
\usepackage{multirow}

\newcommand{\dd}{\mathrm{d}}
\newcommand{\ddCS}{\frac{\dd^2\sigma}{ \dd W \dd Q^2}}
\newcommand{\tdCS}{\frac{\dd^3\sigma}{ \dd W \dd Q^2 \dd \cos\theta^*_\pi}}
\newcommand{\qdCS}{\frac{\dd^4\sigma}{ \dd W \dd Q^2 \dd \Omega_\pi^*}}

\begin{document}

\preprint{APS/123-QED}

\title{Angular distributions in Monte Carlo event generation of weak single-pion production}

\author{K. Niewczas}
\email{kajetan.niewczas@uwr.edu.pl}
\affiliation{Department of Physics and Astronomy, Ghent University, Proeftuinstraat 86, B-9000 Gent, Belgium}
\affiliation{Institute of Theoretical Physics, University of Wroc{\l}aw \\ Plac Maxa Borna 9, 50-204 Wroc{\l}aw, Poland}
\author{A. Nikolakopoulos}
\email{alexis.nikolakopoulos@ugent.be}
\affiliation{Department of Physics and Astronomy, Ghent University, Proeftuinstraat 86, B-9000 Gent, Belgium}
\author{J. T. Sobczyk}
\affiliation{Institute of Theoretical Physics, University of Wroc{\l}aw \\ Plac Maxa Borna 9, 50-204 Wroc{\l}aw, Poland}
\author{N. Jachowicz}
\affiliation{Department of Physics and Astronomy, Ghent University, Proeftuinstraat 86, B-9000 Gent, Belgium}
\author{R. Gonz\'{a}lez-Jim\'{e}nez}
\affiliation{Grupo de F\'isica Nuclear, Departamento de Estructura de la Materia, F\'isica T\'ermica y Electr\'onica, \\
 Universidad Complutense de Madrid and IPARCOS, CEI Moncloa, 28040  Madrid, Spain}
\date{\today}

\begin{abstract}
One of the substantial sources of systematic errors in neutrino oscillation experiments that utilize neutrinos from accelerator sources stems from a lack of precision in modeling single-pion production (SPP).
Oscillation analyses rely on Monte Carlo event generators (MC), providing theoretical predictions of neutrino interactions on nuclear targets.
Pions produced in these processes provide a significant fraction of oscillation signal and background on both elementary scattering and detector simulation levels.
Thus, it is of critical importance to develop techniques that will allow us to accommodate state-of-the-art theoretical models describing SPP into MCs.

In this work, we investigate various algorithms to implement single-pion production models in Monte Carlo event generators.
Based on comparison studies, we propose a novel implementation strategy that combines satisfactory efficiency with high precision in reproducing details of theoretical models predictions, including pion angular distributions.
The proposed implementation is model-independent, thereby providing a framework that can include any model for SPP.
We have tested the new algorithm with the Ghent Low Energy Model for single-pion production implemented in the NuWro Monte Carlo event generator.
\end{abstract}

\maketitle


\section{Introduction}
\label{sec:introduction}

\mbox{Single-pion} production (SPP) is one of the main reaction channels relevant for \mbox{accelerator-based} neutrino experiments, where neutrino energies range from a couple of hundred MeVs up to several GeVs~\cite{NUSTECWP}.
Indeed, in experiments with detectors using Cherenkov radiation, such as T2K~\cite{Abe:2011ks} and MiniBooNE~\cite{Aguilar-Arevalo:2013pmq}, it is challenging to distinguish neutral pions from electrons.
This makes their production the main background for the detection of \mbox{low-energy} electrons.
A good understanding of this background is essential for future CP violation measurements in the Hyper-Kamiokande experiment~\cite{Abe:2015zbg} and in attempts to understand the excess of \mbox{$\nu_e$-like} events reported by the MiniBooNE collaboration~\cite{MB:excessPRL}.
Moreover, produced pions issue a significant background for other neutrino experiments such as MicroBooNE~\cite{MicroBooNE}, as it is challenging to distinguish charged pions from muons in Liquid Argon Time Projection Chambers.
Regarding oscillation analyses, SPP also contributes to the commonly used CC$0\pi$ experimental topology~\cite{Lalakulich:QErec}, provided that the pions get reabsorbed in the nuclear medium or remain otherwise undetected.
Furthermore, this interaction channel is itself a part of the signal for oscillation experiments especially with \mbox{higher-energy} neutrino beams such as NOvA~\cite{NOvA:2018gge} and DUNE~\cite{Abi:2020evt}, but also for T2K~\cite{Abe:2019vii}.

Over the past couple of years, the MINERvA, T2K, ArgoNeuT, and MiniBooNE experiments~\cite{MINERVA:CCPI0, MINERVA:CC1PI, T2KCC1PIH2O, Acciarri:2018ahy, MB:pion, MB:CCneutralpion} have collected an increasingly large dataset for \mbox{(anti-)neutrino-induced} \mbox{single-pion} production on nuclear targets. Subsequently, it has been compared to predictions from several models, revealing significant differences in their description of the data.
Moreover, there are apparent tensions between the MiniBooNE, T2K, and MINERvA SPP measurements~\cite{Sobczyk.:2012zj, HybridRPWIA, Nikolakopoulos:2018gtf, Mosel:2017nzk} themselves.
Ref.~\cite{Stowell:Minervafit2019} showed that a simultaneous agreement between the results of the ANL and BNL bubble chamber data and the MINERvA experiment could not be reached.
Furthermore, it was not possible to provide a consistent description using a single parameter tune for the different SPP channels measured by the latter.

The use of nuclear targets in neutrino oscillation experiments  considerably complicates the description of \mbox{single-pion} production because the presence of such a medium affects all of the hadrons in the process.
On top of that, \mbox{final-state} interactions (FSI), such as pion absorption or charge exchange pion-nucleon scattering, alter the experimental signal entirely.
It is seemingly an intractable problem to provide a detailed microscopic description of FSI over the sizeable phase space of these experiments.
For this reason, the FSI are usually treated in an approximate way using intranuclear cascade models~\cite{NuWroFSI, Niewczas:2019fro, SALCEDO1988} implemented in various Monte Carlo neutrino event generators (MC).
An exception is GiBUU that solves the coupled Boltzmann-Uehling-Uhlenbeck (BUU) transport equations instead~\cite{GiBUUrev}.

A prerequisite for a good description of \mbox{neutrino-induced} \mbox{single-pion} production on nuclei in the factorized approach used in MCs is an accurate model for such scattering off the nucleon.
Several models, with varying regions of applicability, have been developed for \mbox{neutrino-induced} SPP off the nucleon~\cite{ReinSeghal, SL:2003, AmaroResponses, Singh:2006, Hernandez:Pion, BussMosel, Praet, MartiniModel:2009, ZhangSerot, SL, Sato, Singh:2016, SUSA:2016, Gonzalez:SPPnucleon}.
However, these models have not readily found their way into Monte Carlo event generators, and if so, without accounting for their full kinematic complexity.

In this work, we perform a detailed study of possible strategies to implement \mbox{single-pion} production models in neutrino event generators.
Based on the results of this study, we propose a novel algorithm for the case of SPP on the nucleon target to allow for further progress in the accommodation of information from recent experimental measurements.
The algorithm is \mbox{model-independent}, as it only relies on the kinematics of the process and ensures no relevant information is lost on a \mbox{neutrino-nucleon} interaction level.
Such a solution allows for any theoretical model to be implemented in MCs, facilitating a comparison of different approaches.
Additionally, owing to the separation of the leptonic and hadronic currents, it provides flexibility to modify the former, e.g., with Beyond the Standard Model physics.
Furthermore, with appropriately implemented hadronic currents, one can calculate cross sections for charged current, neutral current, and \mbox{electron-induced} SPP with a consistent treatment of both the vector and axial components.
We claim that the proposed algorithm will be of great importance for future implementations of neutrino-nucleus \mbox{single-pion} production, dealing with a considerable number of degrees of freedom and hence a critical demand to maintain both numerical efficiency and precision.

This paper is structured as follows.
In Sec.~\ref{sec:kinematics}, we review the kinematics of \mbox{lepton-induced} \mbox{single-pion} production.
Then, in Sec.~\ref{sec:implementation}, details of the new implementation of SPP in Monte Carlo event generators, as well as the particular numerical tools used, are described.
In Sec.~\ref{sec:results}, we report the results of our study in the context of the implementation performance and physics outcomes.
In the last section, Sec.~\ref{sec:conclusions}, we present our conclusions.

\section{Kinematics and cross section}
\label{sec:kinematics}

\begin{figure}
\includegraphics[width=0.48\textwidth]{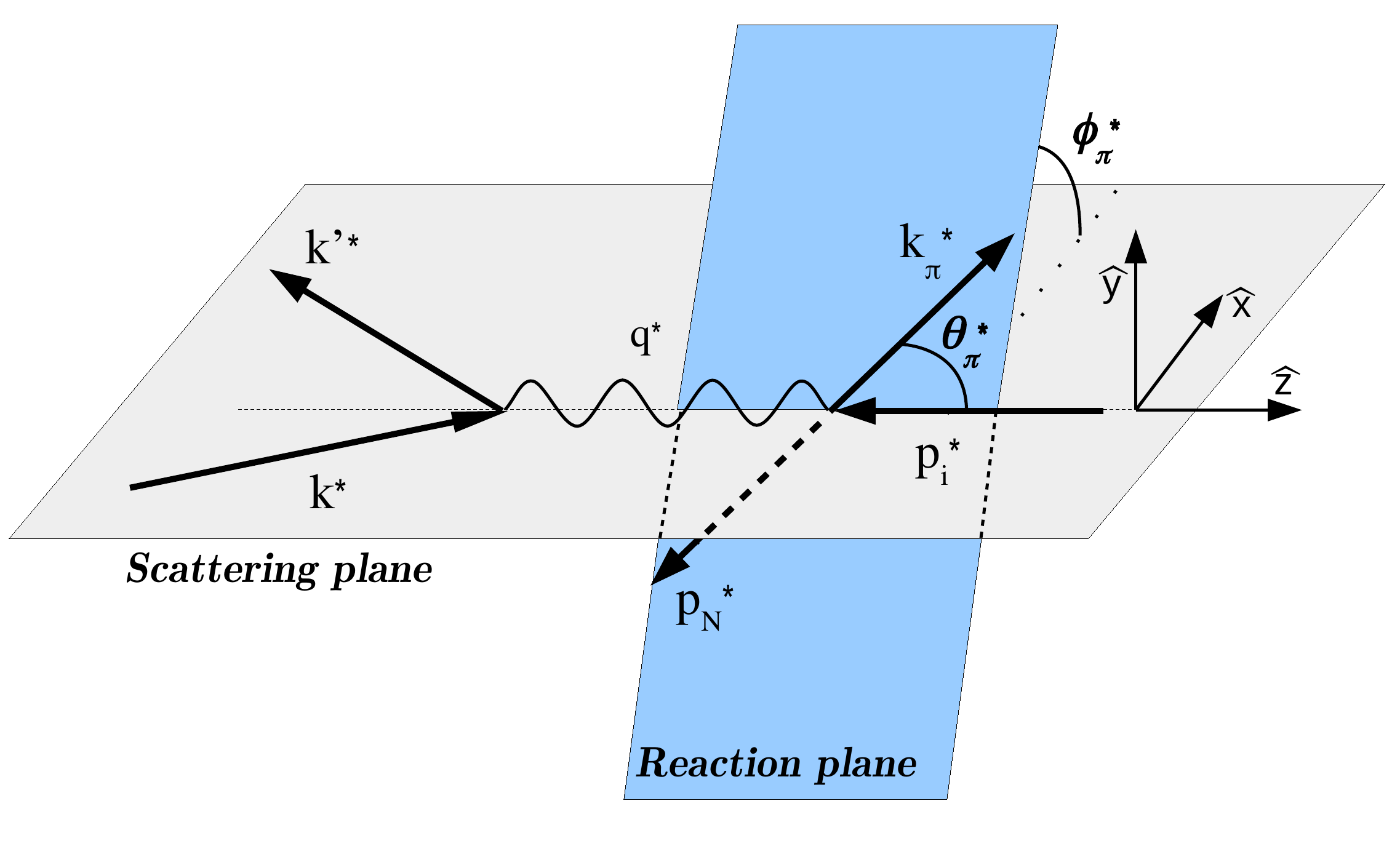}
\caption{Kinematics of lepton-induced single-pion production on the nucleon in the hadronic center-of-momentum frame of reference.}
\label{fig:kinematics}
\end{figure}

We commence by describing the kinematics of \mbox{lepton-induced} \mbox{single-pion} production, where an incoming lepton with \mbox{four-momentum} $k = (E,\vec{k})$ scatters off a nucleon $p_i$ by exchange of a single gauge boson with \mbox{four-momentum} $q = (\omega,\vec{q})$, thereby producing a pion.
We denote the \mbox{four-momenta} of the \mbox{final-state} lepton, pion, and recoiling nucleon by $k^\prime$, $k_\pi$, and $p_N$, and their rest masses by $m$, $M_\pi$, and $M_N$, respectively.
It is convenient to describe such a process in the hadronic \mbox{center-of-momentum} system (CMS), with the lepton plane defining the $x$-$z$ plane and the direction of the momentum transfer $\vec{q}$ defining the $z$-axis, as depicted in Fig.~\ref{fig:kinematics}.
In the hadronic CMS, for which we denote quantities with a superscript ${}^*$, the final hadronic system is at rest, meaning $\vec{k}_\pi^* = -\vec{p}_N^{\:*}$.
We characterize the kinematics by the Lorentz invariants: the invariant hadronic mass $W^2 = (q + p_i)^2 = (k_\pi + p_N)^2$ and the exchanged \mbox{four-momentum} squared $Q^2 = -q^2 = - (k-k^\prime)^2$, along with the produced pion solid angle $\Omega_\pi^*$.

Within the Born approximation, we can describe the cross section as a contraction of the leptonic and hadronic tensors.
The standard calculation of the leptonic tensor for massless incoming leptons yields
\begin{equation}
\label{eq:leptonictensor}
L_{\mu\nu} = k_{\mu}k^{\prime}_{\nu} + k^{\prime}_{\mu}k_{\nu} - \eta_{\mu\nu}k\cdot k^\prime - ih\epsilon_{\mu\nu\alpha\beta}k^{\alpha} k^{\prime\beta},
\end{equation}
where $\eta$ is the metric tensor with signature $(+,-,-,-)$, $\epsilon_{\mu\nu\alpha\beta}$ is the antisymmetric Levi-Civita tensor ($\epsilon_{0123} = +1$), and $h$ is the helicity of the incoming lepton.
We define the hadronic tensor as
\begin{equation}
\label{eq:hadronictensor}
H^{\mu\nu} = \overline{\sum}{J^{\mu}}^\dagger J^\nu,
\end{equation}
with $J^{\mu}$ the hadronic current, and averaging and summation over the spin of the initial and final nucleon are assumed.
Respresenting the hadronic current in terms of initial and final state nucleon spinors and the transition operator $\mathcal{O}^\mu$ as 
\begin{equation}
\label{eq:HadCur}
J^{\mu} = \overline{u}\left(p_N,s_N\right)\mathcal{O}^{\mu}u\left(p_{i}, s_{i}\right),
\end{equation}
one obtains for the hadronic tensor
\begin{equation}
H^{\mu\nu} = \frac{1}{8M^2_N} \mathrm{Tr}\left( \left(\slashed{p}_{i} + M_N \right)       \widetilde{\mathcal{O}}^{\mu} \left(\slashed{p}_N + M_N \right) \mathcal{O}^{\nu} \right),
\end{equation}
where $\widetilde{\mathcal{O}}^{\mu} = \gamma_0\left(\mathcal{O}^\mu\right)^\dagger\gamma_0$.
With these definitions, the cross section is
\begin{equation}
\qdCS = \frac{1}{2}\frac{\mathcal{F}_{CC}^2}{\left(2\pi\right)^4}\frac{|\vec{k}^*_\pi|}{|\vec{k}^{\prime2}|}L_{\mu\nu}H^{\mu\nu},
\end{equation}
where the coupling constant for the charged current case that we consider in this paper is
\begin{equation}
\mathcal{F}_{CC} = \frac{G_F\cos\theta_c}{\sqrt{2}}.
\end{equation}

Using the invariance of the leptonic tensor under rotations of the hadronic plane around $\vec{q}$, one can factorize the dependence of the cross section on the azimuthal angle in terms of trigonometric functions, as shown explicitly in Refs. \cite{Donnelly:1985, Drechsel:1992, SL:2003, Sobczyk:2018}.
Specifically, in the given CMS, we express the cross section as
\begin{equation}
\label{eq:5fold}
\begin{split}
\qdCS &= \frac{1}{2}\frac{\mathcal{F}_{CC}^2}{\left(2\pi\right)^4}\frac{|\vec{k}^*_\pi|}{|\vec{k}^{\prime2}|}\times \left[ A + B\cos\left( \phi_\pi^* \right) \right. \\ 
  &+ \left. C\cos\left( 2\phi_\pi^* \right) + D\sin\left(\phi_\pi^* \right)+ E\sin\left(2\phi_\pi^*\right)   \right],
\end{split}
\end{equation}
where the functions $A$, ..., $E$ do not depend on the azimuthal pion angle $\phi_\pi^*$.
Below, we write them explicitly, in terms of the elements of the leptonic and hadronic tensors, computed for the kinematics of Fig.~\ref{fig:kinematics} (with $\phi_\pi^* = 0$), and making use of the symmetry properties of $L_{\mu\nu}$, as
\begin{equation}
\label{eq:inclusive}
\begin{split}
A &= L_{00}H^{00} + 2L_{30} H_s^{30} + L_{33}H^{33} \\
 &+ \frac{L_{11}+L_{22}}{2}\left(H^{11}+H^{22} \right) + 2iL_{12} H_a^{12},
\end{split}
\end{equation}
\begin{equation}
B = 2L_{01} H_s^{01} + L_{13} H_s^{13} + iL_{02} H_a^{02} + iL_{23} H_a^{23},
\end{equation}
\begin{equation}
C = \frac{L_{11}-L_{22}}{2}\left( H^{11} - H^{22} \right),
\end{equation}
\begin{equation}
D = 2 \left[ -L_{01} H_s^{02} - L_{13}  H_s^{23} + iL_{02} H_a^{01} + iL_{23}  H_a^{13} \right],
\end{equation}
\begin{equation}
E = \left( L_{22} - L_{11} \right) H_s^{12},
\end{equation}
where $H_s$ and $H_a$ correspond to the symmetric (real) and antisymmetric (imaginary) parts of the hadronic tensor:
\begin{equation}
    H^{\mu\nu} = H^{\mu\nu}_s + iH^{\mu\nu}_a, \ \ \ 
    H^{\mu\nu}_{s,a} \in \mathbb{R}.
\end{equation}
For antineutrino interactions, the terms including the imaginary part of the hadronic tensor change sign as all of the off-diagonal terms of the leptonic tensor involving a Lorentz index 2 are purely antisymmetric and proportional to the helicity, while the others are symmetric.

In the context of this work, it is essential to notice that the \mbox{double-differential} cross section $\dd^2\sigma / \dd W \dd Q^2$ and the \mbox{triple-differential} $\dd^3\sigma / \dd W \dd Q^2 \dd \cos\theta_\pi^*$ are entirely determined by the function $A$ as the other contributions disappear after integration over the azimuthal pion angle $\phi_\pi^*$.

We remark that the presented expressions apply to all electroweak SPP processes, thereby facilitating a consistent treatment of the vector current across electron- and \mbox{neutrino-induced} cases implemented in event generators.
Furthermore, a similar separation of the angular dependence is valid for \mbox{one-nucleon} \mbox{knock-out} on a nuclear target or for any \mbox{semi-leptonic} process in which a single \mbox{on-shell} particle defines the hadronic plane for that matter.
Thus, similar methods as the ones outlined in the next section should apply to the implementation of microscopic models for exclusive one-nucleon knock-out.

\section{Monte Carlo event generation for single-pion production}
\label{sec:implementation}

The kinematics for weak \mbox{single-pion} production off the nucleon given an incoming neutrino energy, the target nucleon momentum, and an arbitrarily chosen lepton scattering plane, is fully described by four independent variables.
In what follows, these quantities are considered to be random variables with a probability distribution defined by Eq.~\ref{eq:5fold}.
While constructing Monte Carlo event generators, one of the major tasks is to generate these variables efficiently.
Here, we discuss several of our approaches, each of them presenting a different \mbox{trade-off} between efficiency, precision, and reliance on precomputed assets.

\subsection{4D algorithm}
\label{sec:implementation:4D}

The most straightforward approach is to use directly the full cross section formula~(\ref{eq:5fold}).
The available phase space of the independent variables $W$, $Q^2$, $\cos\theta_\pi^*$, $\phi_\pi^*$ is
\begin{equation}
 \begin{split}
    W \in \ & [M, \sqrt{s}-m], \\
    Q^2 \in \ & [2\underline{E}\underline{E}^\prime-m^2-2\underline{E}|\vec{\underline{k}^\prime}|, 2\underline{E}\underline{E}^\prime-m^2+2\underline
{E}|\vec{\underline{k}^\prime}| ], \\
    \cos\theta_\pi^* \in \ & [-1,1], \\
    \phi_\pi^* \in \ & [0,2\pi],
\end{split}
\label{eq:phasespace}
\end{equation}
where $s=(k+p_i)^2$, and the underline marks the quantities calculated in the lepton+hadron \mbox{center-of-momentum} frame:
\begin{equation}
\begin{split}
    \underline{E}=\frac{s-M_N^2}{2\sqrt{s}},
    \quad
    \underline{E}^\prime=\frac{s+m^2-W^2}{2\sqrt{s}},\\
    |\vec{\underline{k}^\prime}|=\frac{\sqrt{(s-m^2-W^2)^2-4m^2W^2}}{2\sqrt{s}}.
\end{split}
\end{equation}
In this approach, for each event, we sample four independent variables, adding a randomly selected lepton scattering plane.
We perform the sampling following the order presented in Eq.~\ref{eq:phasespace}, starting from $W$, because the range in $Q^2$ is \mbox{$W$-dependent} and because one needs both $W$ and $Q^2$ to specify the hadronic CMS needed to select $\cos\theta_\pi^*$ and $\phi_\pi^*$.
Such information is enough to generate the full kinematics of an event trial.
Each of them has an assigned {\it event weight}, given by Eq.~\ref{eq:5fold} multiplied by the Monte Carlo phase space factor
\begin{equation}
    V_{4D}=(\sqrt{s}-m-M_N)\cdot4\underline{E}|\vec{\underline{k}^\prime}|\cdot 2\cdot 2\pi.
\end{equation}
The average value of this weight is equal to the total cross section.
We obtain the final set of events by applying the \mbox{accept-reject} algorithm on the collection of trials.
We will refer to this strategy of generating events as the "4D algorithm".

Although asymptotically correct, we expect this approach to be inefficient, especially with increasing neutrino energies.
The efficiency of an accept-reject algorithm depends on the interplay between the distribution's shape and the sampling envelope.
Since the former is initially unknown, we choose the latter to be the maximal value of the cross section (Eq.~\ref{eq:5fold}), calculated in real-time.
As we increase the phase space, we access new regions of low, relative to the envelope, cross section values, leading to event trials with a minimal chance of acceptance.
The interplay between the acceptance efficiency and the computation time needed to calculate an event weight are the main features contrasting the proposed algorithms.
In the 4D algorithm, for every event trial, we evaluate the value of the cross section given by Eq.~\ref{eq:5fold} once.

\subsection{3D algorithm}
\label{sec:implementation:3D}

In the second approach, we isolate the dependence of the cross section on the azimuthal pion angle.
After performing an integration over $\phi_\pi^*$, the differential cross section depends only on the function $A$, and its explicit dependence on $H_{\mu\nu}$ reads
\begin{equation}
\begin{split}
    &\tdCS = \frac{1}{2}\frac{\mathcal{F}_{CC}^2}{\left(2\pi\right)^3}\frac{|\vec{k}^*_\pi|}{|\vec{k}^{\prime2}|} \left[  L_{00} H^{00}(W,Q^2,\cos\theta_\pi^*) \right. \\ 
    &+ 2L_{30} H^{30}_s (W,Q^2,\cos\theta_\pi^*)+ \left. L_{33} H^{33}(W,Q^2,\cos\theta_\pi^*)  \right. \\
    &+ \frac{L_{11}+L_{22}}{2} (H^{11} + H^{22})(W,Q^2,\cos\theta_\pi^*) \\
    &+ \left. 2iL_{12} H^{12}_a (W,Q^2,\cos\theta_\pi^*) \right],
\end{split}
\label{eq:responses_3D}
\end{equation}
where the hadronic tensor elements are functions of three variables: $W$, $Q^2$, $\cos\theta_\pi^*$.
In this case, only three variables are sampled and we attribute event trials with weights obtained from multiplying the results of Eq.~\ref{eq:responses_3D} by the new Monte Carlo phase space volume
\begin{equation}
   V_{3D}= (\sqrt{s}-m-M)\cdot4\underline{E}|\vec{\underline{k}^\prime}|\cdot 2.
\end{equation}
As before, the average of the event weights yields the total cross section.
We obtain the final set of, yet incomplete, events using the same accept-reject method, with the sampling envelope given by the maximum of Eq.~\ref{eq:responses_3D}.
Due to the reduced phase space dimensionality, the accept-reject algorithm for incomplete events, without an assigned value of the pion azimuthal angle, is more efficient.

For already selected events, we sample the variable $\phi_\pi^*$ using the known probability distribution given, for fixed values of $W$, $Q^2$, $\cos\theta_\pi^*$, by
\begin{equation}
\begin{split}
  f(\phi_\pi^*) = A
  & + B\cos\left( \phi_\pi^* \right) + C\cos\left( 2\phi_\pi^* \right) \\
  & + D\sin\left(\phi_\pi^* \right)+ E\sin\left(2\phi_\pi^*\right),
\end{split}
\end{equation}
and its cumulative distribution function:
\begin{equation}
\begin{split}
    F(\phi_\pi^*) &= \frac{\phi_\pi^*}{2\pi} + \frac{B}{2\pi A}\sin\phi_\pi^* + \frac{C}{4\pi A}\sin 2\phi_\pi^*\\
    & + \frac{D}{2\pi A}(1-\cos\phi_\pi^*) + \frac{E}{4\pi A}(1-\cos 2\phi_\pi^*).
\end{split}
\end{equation}
As the derivative of the $F(\phi_\pi^*)$ function is known algebraically, its inversion with the Newton method is efficient and converges rapidly. In what follows, we will call this procedure the "3D algorithm".

\subsection{2D algorithm}
\label{sec:implementation:2D}

The starting point for this approach is the formula
\begin{equation}
\label{eq:responses}
\begin{split}
    & \ddCS = \frac{1}{2}\frac{\mathcal{F}_{CC}^2}{\left(2\pi\right)^3}\frac{|\vec{k}^*_\pi|}{|\vec{k}^{\prime2}|} \left[  L_{00} \widetilde{H}^{00}(W,Q^2) \right. \\
    & + 2L_{30} \widetilde{H}^{30}_s (W,Q^2) + L_{33} \widetilde{H}^{33}(W,Q^2) \\
    & + \frac{L_{11}+L_{22}}{2} (\widetilde{H}^{11} + \widetilde{H}^{22}) (W,Q^2) \\
    & + \left. 2iL_{12} \widetilde{H}^{12}_a (W,Q^2) \right],
\end{split}
\end{equation}
obtained from Eq.~\ref{eq:responses_3D} by integrating out the $\cos\theta_\pi^*$ variable and adopting the notation:
\begin{equation}
    \widetilde{H}^{\mu\nu}(W,Q^2) = \int_{-1}^{1} H^{\mu\nu}(W,Q^2,\cos\theta_\pi^*) \mathrm{d\cos\theta_\pi^*}.
\end{equation}
As a result, we express the \mbox{double-differential} cross section in terms of 5 combinations of hadronic tensor elements, which depend solely on $W$ and $Q^2$.
We store their values in the form of lightweight tables.

The first step of the '2D algorithm' is to sample a pair of variables $(W, Q^2)$ with the probability density defined by Eq.~\ref{eq:responses}.
We perform it efficiently, using the precalculated tables with a suitable bilinear interpolation.
At this point, we build an incomplete event trial and compute its weight, analogously to the previous approaches, by multiplying the values obtained from Eq.~\ref{eq:responses} by the Monte Carlo phase space factor
\begin{equation}
   V_{2D}= (\sqrt{s}-m-M)\cdot4\underline{E}|\vec{\underline{k}^\prime}|.
\end{equation}
We accept the set of incomplete events according to their weights, relative to the maximum of Eq.~\ref{eq:responses}, and only then we assign the values of $\cos\theta_\pi^*$ and 
$\phi_\pi^*$.
Such an approach saves a considerable amount of time, avoiding the computation of a full event before applying the \mbox{accept-reject} algorithm.

We select the value of $\cos\theta_\pi^*$ using a probability distribution governed by the function $A$.
To optimize this task, we exploit the smooth character of this function in the region of interest.
Having $W$ and $Q^2$ fixed, we calculate the values of $A(\cos\theta_\pi^*)$ at $k$ points and approximate as a polynomial of degree $k-1$.
Then, we obtain the cumulative distribution function as a polynomial of degree $k$ and sample the $\cos\theta_\pi^*$ variable using the inverse sampling method.
For $(k\leq3)$, we perform the inversion algebraically, while for larger $k$, numerically, with the bisection method.
We have checked that, for most kinematics, the degree of $k=3$ provides sufficient precision, while the distributions are almost exact on the whole phase space for degrees $k\geq7$.
We will discuss the choice of the optimal value of $k$ in the next section.

Depending on the implementation effort and allowed memory, it is also possible to store in tables hadronic tensor elements $H_{\mu\nu}(W,Q^2,\cos\theta_\pi^*)$ that allow obtaining the full $A(W,Q^2,\cos\theta_\pi^*)$ function.
Then, in each event, the maximum of the $\cos\theta_\pi^*$ distribution is given explicitly, and one can sample its value using the \mbox{accept-reject} method.
Such an approach enables us to reduce the \mbox{time-consumption} of each trial event further.
In what follows, we will denote this approach as the "2D algorithm (table)".

Finally, to finish building the kinematics for the accepted events, we need to sample the variable $\phi^*_\pi$.
We proceed by repeating the method described in Sec.~\ref{sec:implementation:3D}.

\subsection{Numerical tools}
\label{sec:implementation:tools}

To reliably test the performance of the abovementioned sampling algorithms, we performed simulations using the Ghent Low Energy Model (LEM) of \mbox{single-pion} production implemented in the NuWro Monte Carlo event generator.
The particular implementation works on a restricted phase space defined by the condition $W < 1.5~\mathrm{GeV}$.

\subsubsection{Ghent Low Energy Model of SPP}

This single-pion production model is based on the work of Hern\'andez, Nieves, and Valverde (HNV), first presented in Ref.~\cite{Hernandez:Pion} with later improvements of Refs.~\cite{Hernandez:PionNucleus, LAR:Watson, Hernandez:2016yfb}.
It contains a microscopic description of the SPP at the amplitude level and includes, in addition to the contributions from the $\Delta$(1232) and $D_{13}$(1520) resonances (both direct and crossed channel Feynman diagrams), the \mbox{lowest-order} background diagrams derived from chiral perturbation theory (ChPT).
Additionally, it includes a relative phase between the ChPT terms and the dominant partial wave of the $\Delta$-pole, which partially restores unitarity~\cite{LAR:Watson}.

The Ghent LEM~\cite{Gonzalez:SPPnucleon} is a custom variant of the model with an independently written code.
On top of the standard version, it includes additional $s$- and $u$-channel contributions from the spin-$^1/_2$ resonances $P_{11}$(1440) and $S_{11}$(1535)~\cite{Lalakulich:Res}.
Additionally, the same model, working in the relativistic plane wave impulse approximation, was extended to describe neutrino scattering on nuclei~\cite{HybridRPWIA,Nikolakopoulos:2018gtf}.

\subsubsection{NuWro Monte Carlo event generator}
\label{sec:implementation:tools:nuwro}

NuWro is a versatile Monte Carlo neutrino event generator, which has been developed by the theoretical group of the University of Wroc{\l}aw since 2005.
It is applicable for simulations in the range of neutrino energies covered by the accelerator-based neutrino oscillation experiments, with an upper bound of $\sim 100$~GeV.
NuWro supports quasielastic, \mbox{single-pion} production, and more inelastic channels (DIS) of neutrino scattering off free nucleons.
The \mbox{neutrino-nucleus} interactions are modeled with various nuclear models (e.g., global or local Fermi gas, spectral functions~\cite{Benhar:1994hw, Ankowski:2014yfa}, or a \mbox{momentum-dependent} nuclear potential~\cite{Juszczak:2005wk}) in the impulse approximation, succeeded by \mbox{final-state} interactions of outgoing hadrons simulated using an intranuclear cascade model~\cite{Niewczas:2019fro, NuWroFSI}.
Moreover, the inclusion of complex nuclear targets enables additional interaction channels such as two-body current processes, coherent pion production, and neutrino scattering off atomic electrons~\cite{Zhuridov:2020hqu}.
The code used in this work bases on NuWro version 19.02.2~\cite{NuWroREPO}.

The NuWro \mbox{single-pion} production model combines the contribution from the $\Delta(1232)$ resonance excitation~\cite{NuWroFF} with a \mbox{non-resonant} background obtained by extrapolating the DIS contribution to lower values of $W$, blended incoherently in the region $W \in (1.3, 1.6)~\mathrm{GeV}$~\cite{Juszczak:2005zs}.
The generated events follow the \mbox{double-differential} cross sections $\dd^2\sigma / \dd W \dd Q^2$ for both the resonant and \mbox{non-resonant} parts.
On top of that, the model obtains the $\Omega_\pi^*$ distributions using the parametrized ones measured by the BNL bubble chamber experiment~\cite{Kitagaki:1986ct} for the former, while for the latter, obtains the kinematics using the PYTHIA6 hadronization routines~\cite{PYTHIA6}.
Alternatively, one can use the parametrization obtained by the ANL experiment~\cite{Radecky:1981fn}.
In this work, we refer to this model as "isobar NuWro".

In NuWro, for all of the described \mbox{single-pion} production model implementations, we apply additional optimizations of sampling in the $(W, Q^2)$ plane.
As it is common for all models and methods presented in this work, this has no impact on our findings nor conclusions.

\section{Results}
\label{sec:results}

\subsection{Performance}
\label{sec:results:performance}

\begin{table*}[ht]
\centering
\begin{subtable}{0.49\linewidth} 
\resizebox{\linewidth}{!}{%
\begin{tabular}{c c c c c c c c}
\hline \hline \multicolumn{2}{c}{model} & $\sigma [\mathrm{cm}^2]$ & $s_\mathrm{1M} [\mathrm{cm}^2]$ & $\tau$ & $\epsilon$ & $\alpha$ & $S_\mathrm{1M}$ \\ \hline 
\multicolumn{2}{c}{4D alg.} & 5.1724e-39 & 7.8e-42 & 8.01e-07 & 0.12 & -  & 6.9 \\ 
\multicolumn{2}{c}{3D alg.} & 5.1661e-39 & 7.7e-42 & 8.02e-07 & 0.13 & 1.0 & 6.9 \\ 
\multirow{3}{*}{\rotatebox[origin=c]{90}{2D alg.}} & $(k=7)$ & 5.1586e-39 & 7.5e-42 & 4.04e-08 & 0.16 & 143.9 & 6.1 \\ 
 & $(k=3)$ & 5.1623e-39 & 7.5e-42 & 4.04e-08 & 0.16 & 72.0 & 3.2 \\ 
 & (table) & 5.1613e-39 & 7.5e-42 & 4.03e-08 & 0.16 & 18.6 & 1.0 \\ \hline \hline 
\end{tabular}}%
\caption{$E = 1.0 \ \mathrm{GeV}$ neutrinos off proton target.}
\end{subtable}%
\hfill \vspace{0.1cm} 
\begin{subtable}{0.49\linewidth} 
\resizebox{\linewidth}{!}{%
\begin{tabular}{c c c c c c c c}
\hline \hline \multicolumn{2}{c}{model} & $\sigma [\mathrm{cm}^2]$ & $s_\mathrm{1M} [\mathrm{cm}^2]$ & $\tau$ & $\epsilon$ & $\alpha$ & $S_\mathrm{1M}$ \\ \hline 
\multicolumn{2}{c}{4D alg.} & 2.5105e-39 & 2.7e-42 & 1.83e-06 & 0.15 & -  & 12.1 \\ 
\multicolumn{2}{c}{3D alg.} & 2.5095e-39 & 2.7e-42 & 1.83e-06 & 0.18 & 0.5 & 11.2 \\ 
\multirow{3}{*}{\rotatebox[origin=c]{90}{2D alg.}} & $(k=7)$ & 2.5126e-39 & 2.6e-42 & 4.11e-08 & 0.21 & 169.4 & 7.2 \\ 
 & $(k=3)$ & 2.5124e-39 & 2.6e-42 & 4.10e-08 & 0.21 & 85.1 & 3.7 \\ 
 & (table) & 2.5116e-39 & 2.6e-42 & 4.08e-08 & 0.21 & 22.0 & 1.1 \\ \hline \hline 
\end{tabular}}%
\caption{$E = 1.0 \ \mathrm{GeV}$ neutrinos off neutron target.}
\end{subtable}%
\hfill \vspace{0.1cm} 
\begin{subtable}{0.49\linewidth} 
\resizebox{\linewidth}{!}{%
\begin{tabular}{c c c c c c c c}
\hline \hline \multicolumn{2}{c}{model} & $\sigma [\mathrm{cm}^2]$ & $s_\mathrm{1M} [\mathrm{cm}^2]$ & $\tau$ & $\epsilon$ & $\alpha$ & $S_\mathrm{1M}$ \\ \hline 
\multicolumn{2}{c}{4D alg.} & 6.8637e-39 & 11.2e-42 & 8.04e-07 & 0.08 & -  & 9.9 \\ 
\multicolumn{2}{c}{3D alg.} & 6.8634e-39 & 10.8e-42 & 8.01e-07 & 0.10 & 1.0 & 8.8 \\ 
\multirow{3}{*}{\rotatebox[origin=c]{90}{2D alg.}} & $(k=7)$ & 6.8327e-39 & 10.5e-42 & 3.98e-08 & 0.12 & 149.1 & 6.3 \\ 
 & $(k=3)$ & 6.8510e-39 & 10.5e-42 & 4.08e-08 & 0.12 & 72.6 & 3.3 \\ 
 & (table) & 6.8450e-39 & 10.5e-42 & 4.04e-08 & 0.12 & 19.0 & 1.1 \\ \hline \hline 
\end{tabular}}%
\caption{$E = 2.5 \ \mathrm{GeV}$ neutrinos off proton target.}
\end{subtable}%
\hfill 
\begin{subtable}{0.49\linewidth} 
\resizebox{\linewidth}{!}{%
\begin{tabular}{c c c c c c c c}
\hline \hline \multicolumn{2}{c}{model} & $\sigma [\mathrm{cm}^2]$ & $s_\mathrm{1M} [\mathrm{cm}^2]$ & $\tau$ & $\epsilon$ & $\alpha$ & $S_\mathrm{1M}$ \\ \hline 
\multicolumn{2}{c}{4D alg.} & 4.5860e-39 & 4.7e-42 & 1.84e-06 & 0.14 & -  & 13.5 \\ 
\multicolumn{2}{c}{3D alg.} & 4.5851e-39 & 4.4e-42 & 1.83e-06 & 0.18 & 0.5 & 11.4 \\ 
\multirow{3}{*}{\rotatebox[origin=c]{90}{2D alg.}} & $(k=7)$ & 4.5762e-39 & 4.2e-42 & 4.19e-08 & 0.20 & 169.6 & 7.3 \\ 
 & $(k=3)$ & 4.5805e-39 & 4.2e-42 & 4.13e-08 & 0.20 & 86.0 & 3.8 \\ 
 & (table) & 4.5809e-39 & 4.2e-42 & 4.12e-08 & 0.20 & 22.3 & 1.1 \\ \hline \hline 
\end{tabular}}%
\caption{$E = 2.5 \ \mathrm{GeV}$ neutrinos off neutron target.}
\end{subtable}%
\hfill 
\caption{Tables of the performance of the algorithms, based on 1M event simulations. The values of $\tau$ are normalized to obtain $S_\mathrm{1M} = 1.0$ for the "2D alg. (table)" model.}
\label{tab:performance}
\end{table*}

\begin{figure*}[ht]
\includegraphics[width=\textwidth]{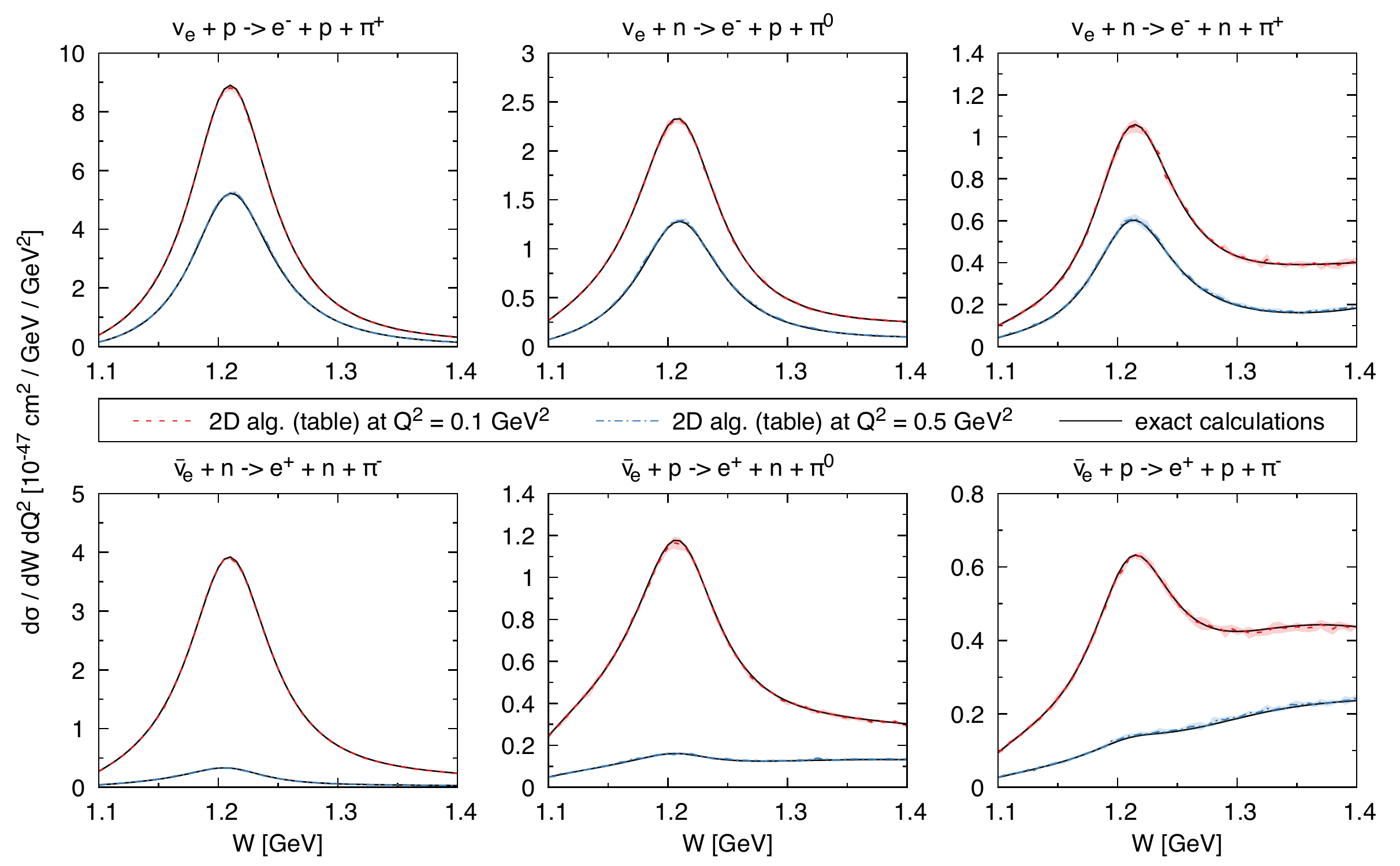}
\caption{Double-differential cross sections for the $\nu_e$- and $\overline{\nu}_e$-induced single-pion production processes as a function of $W$ for different values of $Q^2$ with incoming energy $E = 1~\mathrm{GeV}$. Solid lines are showing the Ghent LEM results, while the (dot-)dashed ones are results of the "2D algorithm (tables)" method implemented in NuWro.}
\label{fig:ddcs_WQ2_fixQ2}
\end{figure*}

We have implemented the Ghent LEM in NuWro, applying the five versions of the strategies presented in Sec.~\ref{sec:implementation}, labeled: "4D alg.", "3D alg.", "2D alg. ($k=3$)", "2D alg. ($k=7$)", "2D alg.
(table)".
We summarize their performance in Table~\ref{tab:performance}, with four numerical computations: for two neutrino energies $E = 1.0, \ 2.5~\mathrm{GeV}$, and off both proton and neutron nucleon targets.
In the respective columns of these tables, one can find: an average weight $\sigma$ and its standard deviation $s_\mathrm{1M}$ calculated from 1 million trial events, computer time $\tau$ needed to calculate a trial event (before the accept-reject algorithm is applied) in arbitrary units, efficiency $\epsilon$ of the accept-reject algorithm, and the relative increase of computer time $\alpha$ needed to generate an event with complete kinematics.
In the last columns, we present values of $S_\mathrm{1M}$ that is a measure of the performance of a given algorithm: an estimate of the time needed to produce a sample of $N=1\times10^6$ events.
In a given simulation of efficiency $\epsilon$, one has to generate $N/\epsilon$ trial events, out of which $N$ events are accepted and require the complete kinematics, while $N/\epsilon - N$ are the rejected trial events that require only the weight calculation.
Since the computation of a trial event takes time $\tau$ and of a complete event $\tau(1+\alpha)$, the overall computer time needed to generate a set of $N$ events becomes
\begin{equation}
    S_N=N\cdot\tau\cdot(1+\alpha)+(\frac{N}{\epsilon}-N)\cdot\tau=N\cdot\tau\cdot (\frac{1}{\epsilon}+\alpha).
\end{equation}
Thus, the value of $S_N$ depends on three variables: $\tau$, $\epsilon$, $\alpha$, that fully characterize each algorithm.

The first, most significant difference between various approaches appears in the values of $\tau$ and show that, in the models used, the time needed for generating a trial event is $\sim20$ times smaller for the 2D algorithm off protons, relative to the 3D and 4D algorithms, while off neutrons the difference rises about twice as much.
The former stems solely from the computational cost needed to evaluate the hadronic tensor, which is the bottleneck of the Ghent LEM, while the latter comes from the fact that \mbox{neutrino-induced} SPP off the neutron involves two possible final states and both cross sections need to be evaluated to obtain the weight of any of those events.

The differences in $\epsilon$ for simulations with the same conditions come from the differential cross section shapes as well as the size of the sampled phase space that grows with increasing energy.
Due to the dominance of the $\Delta^{++}$ resonance, the SPP cross sections for neutrino scattering off the proton target are much more peaked, leading to lower event acceptance efficiency.
On the other hand, the differences in efficiencies between particular algorithms within the same simulations come from different dimensionalities of the sampled phase spaces and the fact that the cross sections are not uniform in the additional variables ($\cos\theta_\pi^*$, $\phi_\pi^*$).

Values of the third characteristic variable $\alpha$ represent all of the secondary effort, relative to the event trial computation time, needed to generate the full kinematics of an accepted event.
One can see that for the 3D algorithm, in which sampling of the $\phi_\pi^*$ variable requires to compute the hadronic tensor one additional time, relative to the 4D method, $\alpha$ equals $1.0$ and $0.5$ for the proton and neutron targets, respectively.
The 2D algorithm methods, on top of the $\phi_\pi^*$ sampling, require additional effort to assign the $\cos\theta_\pi^*$ variable.
The increase in $\alpha$ while going from the 2D (table) method to the ones that use polynomial interpolation is almost proportional to the number of times $(k=3, 7, ...)$ we calculate the hadronic tensor.
We expect that one can avoid such behavior using a model implementation that separates the angular dependence algebraically, e.g., in a partial wave expansion, where one can compute the hadronic tensor for different values of $\cos\theta_\pi^*$ at fixed values of $Q^2$ and $W$ in a much shorter time.
However, in general, the $\cos\theta_\pi^*$ dependence is not a priori known.
Thus, in this study, we opted to present the most \mbox{model-independent} case.

The resultant performance of all the optimization methods in reducing the total simulation time $S_N$ is notable.
Considering its execution time and susceptibility to the investigated factors, we conclude that the "2D alg. (table)" method performs best, and in what follows, we use it to generate all of the Monte Carlo simulation results.
To strengthen this reasoning, we emphasize that in actual simulations, there is an additional, global computational effort needed to specify the weight of particular interaction channels and initialize the event sampling envelope.
In NuWro, we know it as generating {\it test events} that require solely an event weight calculation, which is less demanding using the 2D algorithms.

\subsection{Inclusive cross section}
\label{sec:results:inclusive}

To illustrate the accuracy of the implementation of the \mbox{single-pion} production model in NuWro within the "2D alg. (table)" framework, we show several comparisons with the exact results obtained with the original Ghent LEM code.
For every presented plot, we compute the Monte Carlo results by averaging over six simulations with 10M events across the whole phase space.
The additional band represents a $1\sigma$ error on the average.

In Fig.~\ref{fig:ddcs_WQ2_fixQ2}, we compare the results for the inclusive cross sections as a function of $W$ at fixed $Q^2$ for electron neutrinos and antineutrinos with an energy of $E = 1~\mathrm{GeV}$, including all possible \mbox{single-pion} production channels.
For each value of $Q^2 = 0.1, \ 0.5~\mathrm{GeV}^2$, we gathered Monte Carlo events in bins with a width of $\Delta Q^2 = 0.01~\mathrm{GeV}^2$ and $\Delta W = 5~\mathrm{MeV}$.
One can see that the "2D alg. (table)" method provides excellent accuracy.
The statistical uncertainty on its results is the smallest for (anti)neutrino reactions on the (neutron)proton, as these are cases with only a single SPP channel accessible.
For the other target/helicity combinations, the simulations split the events over two final states, with the one of the higher cross section receiving a larger share, which is reflected in the uncertainty.

\subsection{Angular distributions of the pion}
\label{sec:results:angles}

\begin{figure*}[ht]
\includegraphics[width=\textwidth]{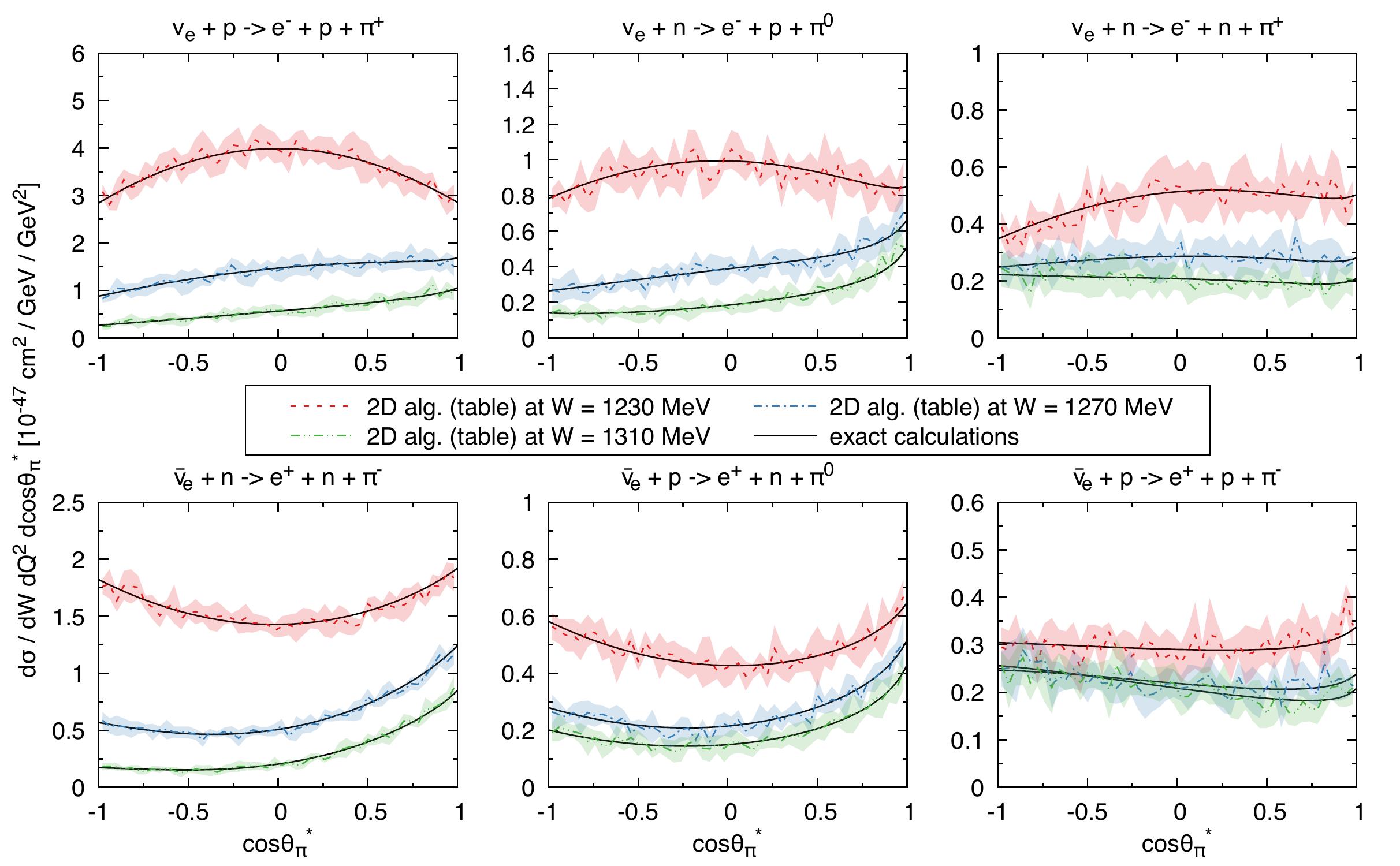}
\caption{Triple-differential cross sections for the $\nu_e$- and $\overline{\nu}_e$-induced single-pion production processes as a function of $\cos\theta_\pi^*$ for different values of $W$ and fixed $Q^2=0.1~\mathrm{GeV}^2$ with incoming energy $E = 1~\mathrm{GeV}$. Solid lines show the Ghent LEM results, while the (dot-)dashed ones are results of the "2D algorithm (tables)" method implemented in NuWro.}
\label{fig:tdcs_WQ2costh_fixQ2W}
\end{figure*}

The main strength of the presented approach is the exact implementation of the outgoing pion angular distributions.
To illustrate this, in Fig.~\ref{fig:tdcs_WQ2costh_fixQ2W}, we plot the cross sections as a function of $\cos\theta^*_\pi$ for values of $W = 1230, \ 1270, \ 1310~\mathrm{MeV}$ and fixed $Q^2=0.1~\mathrm{GeV}^2$ with incoming (anti)neutrino energy $E=1~\mathrm{GeV}$.
We obtained NuWro results in the same way as described in Sec.~\ref{sec:results:inclusive}.
Here, we gathered events in bins of $\Delta Q^2 = 0.01~\mathrm{GeV}^2$, $\Delta W = 5~\mathrm{MeV}$, and $\Delta\cos\theta_\pi^* = 0.04$.
The obtained Monte Carlo results precisely reproduce the exact model calculations.
The shape of the $\cos\theta_\pi^*$ distribution varies with both the interaction channel and kinematics.
This behavior is in contrast to the commonly used approach in which the angular dependence of the outgoing \mbox{pion-nucleon} pair is described isotropically or by a distribution independent of the kinematics.

\begin{figure*}[ht]
\includegraphics[width=\textwidth]{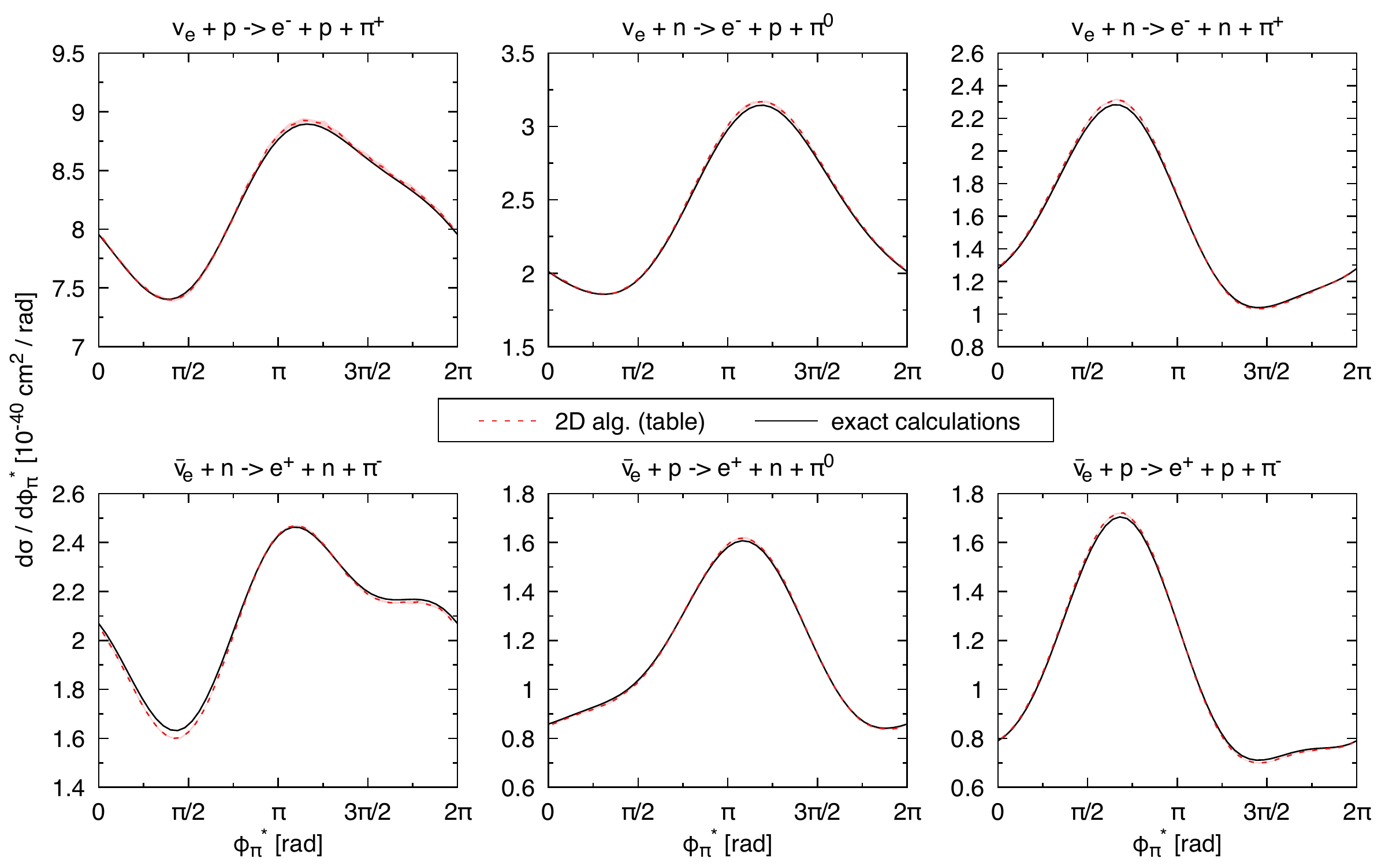}
\caption{Single-differential cross sections for the $\nu_e$- and $\overline{\nu}_e$-induced single-pion production processes as a function of $\phi_\pi^*$ with incoming energy $E = 1~\mathrm{GeV}$. Solid lines show the Ghent LEM results, while the dashed ones are results of the "2D algorithm (tables)" method implemented in NuWro.}
\label{fig:sdcs_phi}
\end{figure*}

The next comparison, in Fig.~\ref{fig:sdcs_phi}, concerns the \mbox{single-differential} cross sections as a function of $\phi_\pi^*$ for electron neutrinos and antineutrinos with an energy of $E = 1~\mathrm{GeV}$, averaged over $\Delta\phi_\pi^* = \pi/25~\mathrm{rad}$ bins.
Since the procedure for sampling $\phi_\pi^*$ is practically exact, shapes of these distributions exemplify the total numerical error propagating from the bilinear and trilinear interpolation of the tabularized information used to sample the values of ($W$, $Q^2$) and $\cos\theta_\pi^*$, respectively.
Hence, one can interpret this comparison as a good measure of the full accuracy of the proposed algorithm.
Regarding the physical results themselves, one immediately notices the asymmetry of $\dd\sigma / \dd\phi_\pi^*$ around $\phi_\pi^*=\pi$, corresponding to pions produced above or below the lepton scattering plane.
As seen in Eq.~\ref{eq:5fold}, the $D$ and $E$ functions, which give contributions proportional to $\sin\left(\phi_\pi^*\right)$ and $\sin\left(2\phi_\pi^*\right)$, respectively, are responsible for such behavior.
As explained thoroughly in Ref.~\cite{Sobczyk:2018}, these asymmetries emerge from relative phase differences between the distinct contributions to the amplitude.
Hence, they are not present in models that are described by incoherent sums of resonances, or resonance and background contributions.
In the Ghent Low Energy Model, the asymmetry can only arise from the interference between the imaginary part of the resonance propagator (plus the Olsson phases in the case of $\Delta$) and the \mbox{non-resonant} background.
Such an asymmetry is also not present in unpolarized electron scattering because both the $D$ function, with the \mbox{vector-vector} contribution proportional to the polarization, and the $E$ function, being a purely \mbox{vector-axial} interference term, disappear in that case.

\begin{figure*}[ht]
\includegraphics[width=\textwidth]{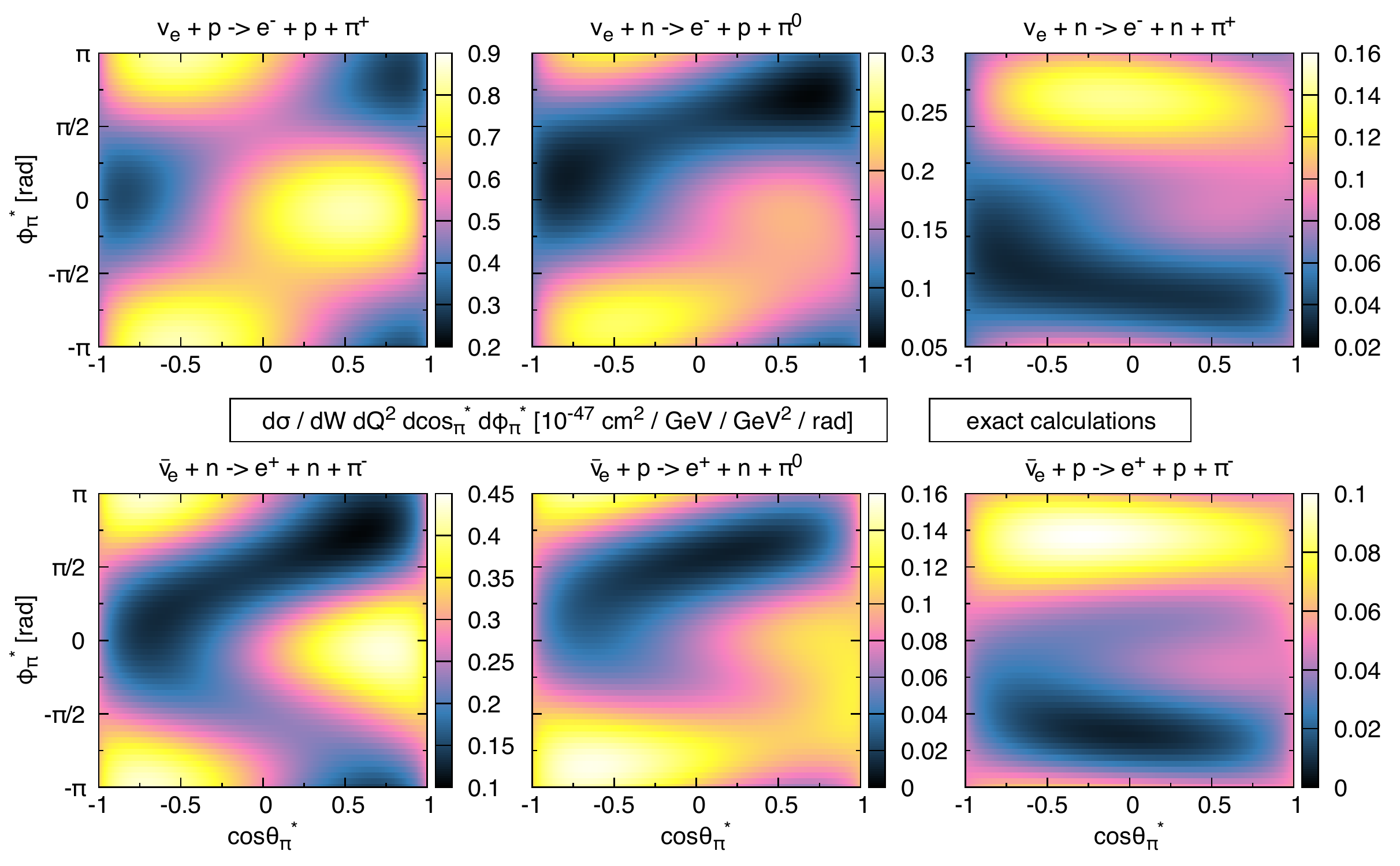}
\caption{Quadruple-differential cross sections for the $\nu_e$- and $\overline{\nu}_e$-induced single-pion production processes as a function of $\cos\theta_\pi^*$ and $\phi_\pi^*$ for fixed $W = 1230~\mathrm{MeV}$ and $Q^2 = 0.1~\mathrm{GeV}^2$ with incoming energy $E = 1~\mathrm{GeV}$. The presented heatmaps are the Ghent LEM results.}
\label{fig:qdcs_WQ2Om_fixQ2W_gent}
\end{figure*}

\begin{figure*}[ht]
\includegraphics[width=\textwidth]{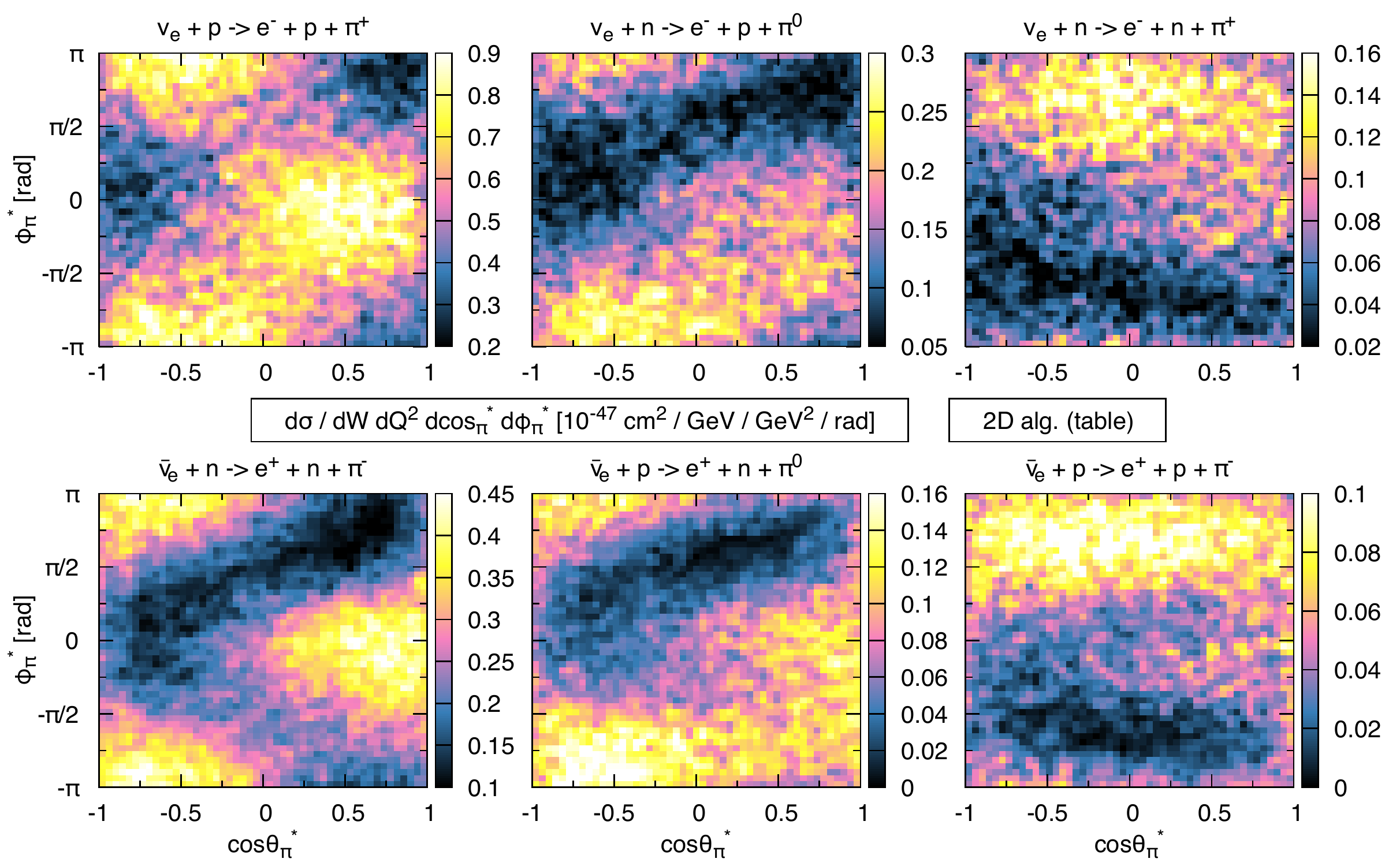}
\caption{Quadruple-differential cross sections for the $\nu_e$- and $\overline{\nu}_e$-induced single-pion production processes as a function of $\cos\theta_\pi^*$ and $\phi_\pi^*$ for fixed $W = 1230~\mathrm{MeV}$ and $Q^2 = 0.1~\mathrm{GeV}^2$ with incoming energy $E = 1~\mathrm{GeV}$. The presented heatmaps are the results of the "2D algorithm (tables)" method implemented in NuWro.}
\label{fig:qdcs_WQ2Om_fixQ2W_2d_tab}
\end{figure*}

\begin{figure*}[ht]
\includegraphics[width=\textwidth]{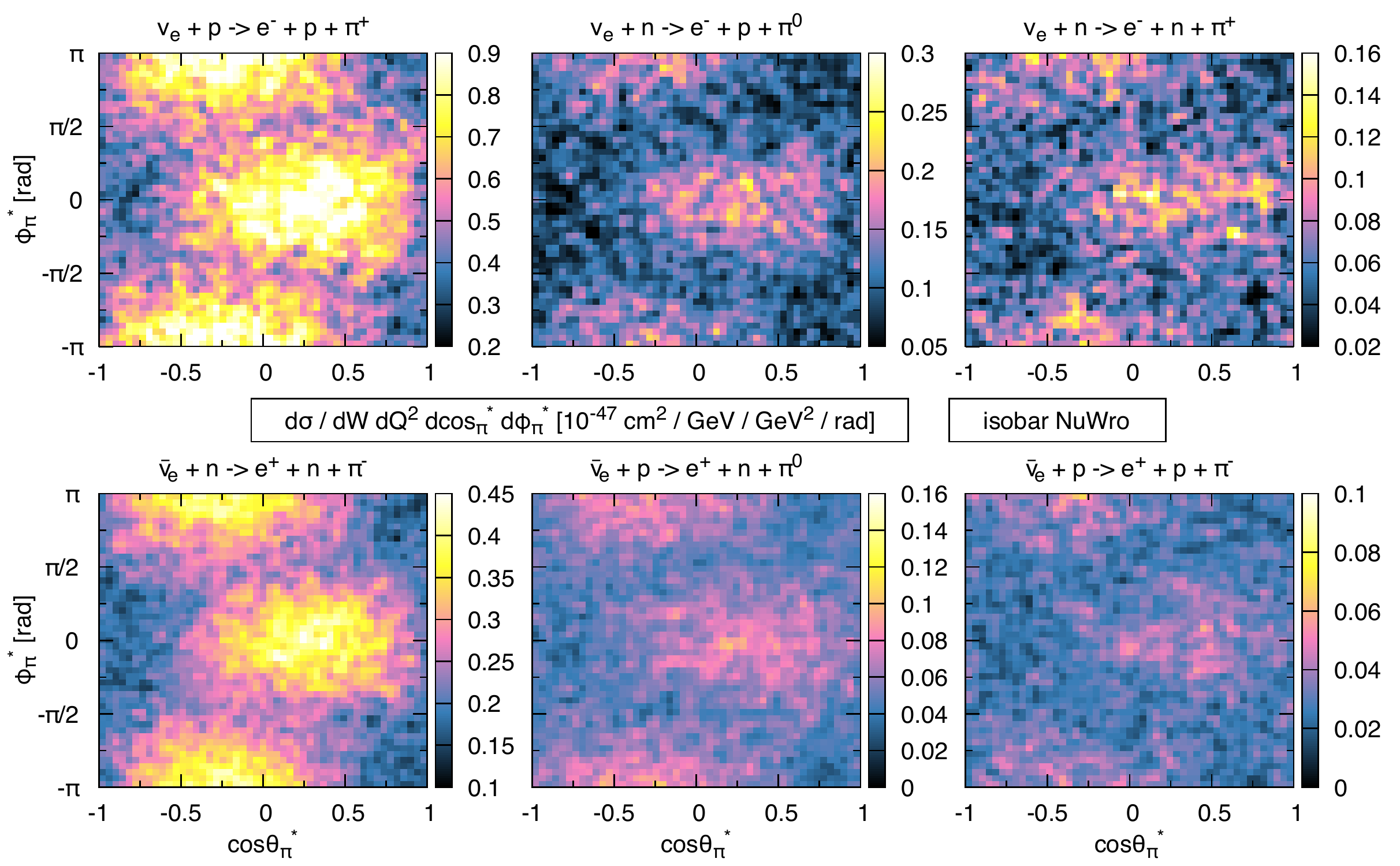}
\caption{Quadruple-differential cross sections for the $\nu_e$- and $\overline{\nu}_e$-induced single-pion production processes as a function of $\cos\theta_\pi^*$ and $\phi_\pi^*$ for fixed $W = 1230~\mathrm{MeV}$ and $Q^2 = 0.1~\mathrm{GeV}^2$ with incoming energy $E = 1~\mathrm{GeV}$. The presented heatmaps are the results of the "isobar NuWro" model, the nominal single-pion production choice.}
\label{fig:qdcs_WQ2Om_fixQ2W_adler}
\end{figure*}

In Figs.~\ref{fig:qdcs_WQ2Om_fixQ2W_gent} and~\ref{fig:qdcs_WQ2Om_fixQ2W_2d_tab}, we show the full \mbox{two-dimensional} $\Omega_\pi^*$ dependence in the different electron \mbox{(anti)neutrino-induced} SPP channels with $E = 1~\mathrm{GeV}$ for fixed $Q^2 = 0.1~\mathrm{GeV}^2$ and $W = 1230~\mathrm{MeV}$, i.e., at the $\Delta(1232)$ peak.
Here, we average the Monte Carlo results over $\Delta Q^2 = 0.01~\mathrm{GeV}^2$, $\Delta W = 5~\mathrm{MeV}$, $\Delta\cos\theta_\pi^* = 0.04$, and $\Delta\phi_\pi^* = \pi/25~\mathrm{rad}$ bins.
Although we performed these NuWro simulations again in the same way as described in Sec.~\ref{sec:results:inclusive}, it is challenging to produce a sufficiently large sample of events to reduce statistical fluctuations.
Still, the agreement we find is remarkably good.
Analyzing the presented distributions, one can see that the $\nu (p,p\pi^+)$ and $\overline{\nu}(n,n\pi^-)$ cross sections are roughly symmetric with respect to $\phi_\pi^*$.
These interaction channels only allow isospin $3/2$ contributions in the \mbox{s-channel} and are thus dominated by the $\Delta(1232)$ resonance, with minimal impact from the background and thereby minimal interference to generate the asymmetry.
The other channels, however, do show a more asymmetric shape as the background contribution grows in relative importance.

In Fig.~\ref{fig:qdcs_WQ2Om_fixQ2W_adler}, we also present the $\Omega_\pi^*$ dependence of the "isobar NuWro" model for the same kinematical setup.
This model uses angular distributions from the BNL parametrization of Ref.~\cite{Kitagaki:1986ct}, as implemented in NuWro 19.02.2.
In the case of neutrino-induced charged pion production off the proton these results are similar to the Ghent LEM, while for the other reaction channels they are quite different.
Such behavior originates from the fact that the BNL (ANL) parametrization is obtained from data for the former reaction in the $\Delta(1232)$ region.
This comparison illustrates that a straightforward application of the same angular distribution to other reaction channels and other phase space regions should be avoided.

\begin{figure*}[ht]
\includegraphics[width=\textwidth]{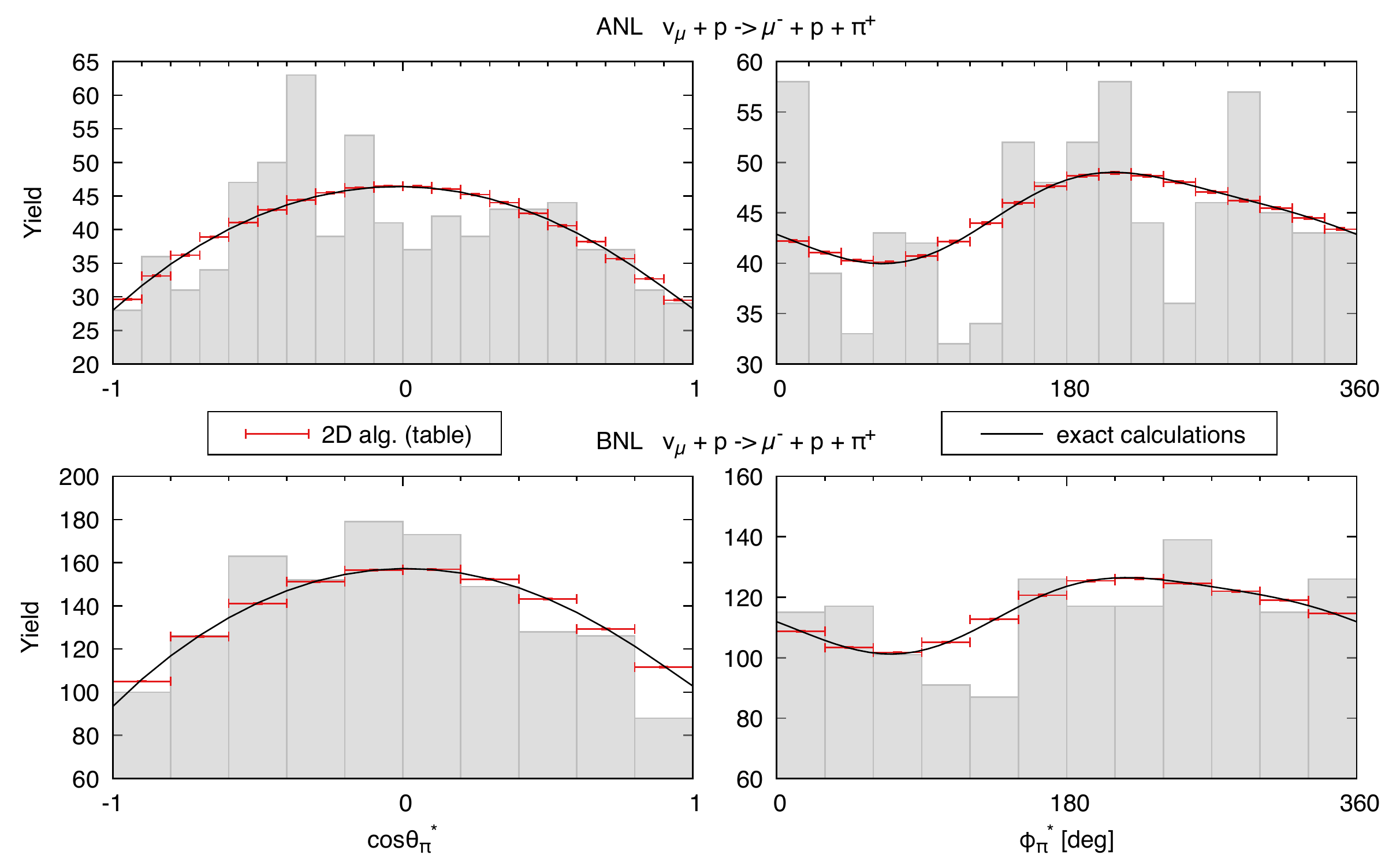}
\caption{Pion angular distributions for the neutrino-induced single-pion production on the proton as a function of $\cos\theta_\pi^*$ or $\phi_\pi^*$, and data from the ANL~\cite{Radecky:1981fn} and BNL~\cite{Kitagaki:1986ct} bubble chamber experiments. Solid lines show the Ghent LEM results, while the bins are the results of the "2D algorithm (tables)" method implemented in NuWro. We normalize our cross section predictions to the total experimental yield.}
\label{fig:sdcs_xNL}
\end{figure*}

Finally, in Fig.~\ref{fig:sdcs_xNL}, we show a \mbox{shape-only} comparison with the \mbox{single-differential} cross sections $\dd\sigma/\dd\cos\theta_\pi^*$ and $\dd\sigma/\dd\phi_\pi^*$ measured by the ANL~\cite{Radecky:1981fn} and BNL~\cite{Kitagaki:1986ct} experiments.
To obtain this, we performed simulations with the $E_\mathrm{ANL}\in(0.2, 6.1)~\mathrm{GeV}$ and $E_\mathrm{BNL}\in(0.1, 7.5)~\mathrm{GeV}$ muon neutrino fluxes off a proton target, and applied a cut on the invariant hadronic mass $W<1.4~\mathrm{GeV}$.
The theoretical results provide a good agreement for the $\dd\sigma/\dd\cos\theta_\pi^*$ differential cross section, especially in the BNL case, while due to the lack of statistics, the $\dd\sigma/\dd\phi_\pi^*$ results are not conclusive.

\section{Conclusions}
\label{sec:conclusions}

The upcoming precision era of neutrino oscillation experiments requires significant improvements in detecting and modeling more exclusive observables of neutrino scattering.
In the case of pion production, a pressing issue is that of their angular distributions.
Due to the lack of available data to constrain these quantities, it is of great importance to equip Monte Carlo neutrino event generators with predictions of the most sophisticated theoretical models available.
We have made an important step towards this goal, focusing on the \mbox{(anti)neutrino-induced} \mbox{single-pion} production on the nucleon and implementing the Ghent Low Energy Model of Ref.~\cite{Gonzalez:SPPnucleon} into the NuWro Monte Carlo event generator.

To this end, we investigated various general, \mbox{model-independent} and efficient implementations based on the separation of the pion angular dependence in the hadronic \mbox{center-of-momentum} reference frame.
They originate from the idea to sample particular independent kinematic variables needed to build a Monte Carlo event in a specific order, using increasingly differential cross section formulas.
In the consecutive steps, such an approach allows performing the \mbox{time-consuming} microscopic model computations only for accepted events and exploits differences in Monte Carlo event generation efficiencies.
All of our algorithms start from sampling the ($W$, $Q^2$) phase space to be able to specify the hadronic CMS.
Then, we use approximations that allow us to efficiently choose the value of $\cos\theta_\pi^*$ as well as to exploit the algebraic dependence of the cross section on $\phi_\pi^*$.
Such approaches provide the flexibility of choosing a different \mbox{trade-off} between efficiency, precision, and reliance on precomputed assets.

To quantify the performance of our implementations, we performed various simulations and measured characteristic quantities for each solution.
We conclude that the method labeled "2D alg. (table)", that exploits all the optimizations and mostly relies on precomputed assets, is the most effective solution.
We exhaustively checked its accuracy in investigating different \mbox{multiple-differential} cross sections, each time obtaining excellent agreement.
The performance of the "2D alg. ($k\geq3$)" methods, which employ a polynomial fit of degree $k-1$ for the $\cos\theta_\pi^*$ probability distributions, was intermediate and strongly relied on the choice of the  degree $k$.
Although we obtained promising results in the $\Delta$-region for $k$-values as low as 3,  it was necessary to perform simulations with degree $k\geq7$ to reproduce the Ghent LEM predictions over the entire investigated phase space to the percent level.
Still, such an approach proved useful, as such a high level of accuracy is not necessary for \mbox{flux-averaged} distributions, and it does not require to precompute nor store the additional tables.

Finally, we compared the new implementation of the Ghent LEM model with the SPP angular distribution data from ANL and BNL bubble chamber experiments and with the results of the nominal \mbox{single-pion} production model of NuWro.
We concluded that it is not feasible to use experimental parametrizations for \mbox{neutrino-induced} SPP off the proton in the $\Delta(1232)$ region for all channels across the whole phase space.
It is of great importance to design Monte Carlo event generators able to provide reliable predictions for such observables.

This work facilitates further studies of nuclear effects in SPP as we can implement more sophisticated models in NuWro.
The next important step of this research will be the extension of this implementation framework to \mbox{single-pion} production on the nucleus and investigation of the precision of various theoretical assumptions of modeling SPP in Monte Carlo event generators.

\begin{acknowledgments}
  This work was supported by the Interuniversity Attraction Poles Programme initiated by the Belgian Science Policy Office (BriX network P7/12) and the Research Foundation Flanders (FWO-Flanders), and by the Special Research Fund, Ghent University.
  KN and JS acknowledge support provided by the NCN Opus Grant No. 2016/21/B/ST2/01092 and also by the Polish Ministry of Science and Higher Education, Grant No. DIR/WK/2017/05.
  RGJ was partially supported by Comunidad de Madrid and UCM under the contract No. 2017-T2/TIC-5252.
\end{acknowledgments}

\bibliographystyle{apsrev4-1.bst}
\bibliography{Bibliography}

\end{document}